# A Classical Key-Distribution System based on Johnson (like) noise – How Secure?

Jacob Scheuer
*Center for the Physics of Information, California Inst. of Technology, Pasadena, CA*

Amnon Yariv
*Department of Applied Physics, California Inst. of Technology, Pasadena, CA*

**Abstract**: We present a comprehensive analysis of the Johnson (like) noise based classical key-distribution scheme presented by Kish [1]. We suggest two passive attack strategies that enable an adversary to gain complete knowledge of the exchanged key. The first approach exploits the transient response of the voltage in the transmission line after the resistors are switched and the second one exploits the finite impedance of the wire connecting the two parties.

## 1. Introduction

Highly secure communication channels are essential elements for numerous present and contemplated applications. Classical encryption schemes utilizing one-way functions offer only computationally hard security, and therefore, their security can be compromised, at least in principle.

Quantum key-distribution systems (QKDS) represent a different approach enabling, in principle, unconditionally secure key-distribution based on the laws of quantum mechanics. Although completely secure, QKDS pose major technological challenges which limit significantly the key-establishing rates (<100kHz) and the achievable ranges (<120km).

Recently, a classical KDS scheme utilizing Johnson noise in resistors was suggested by Kish [1]. A schematic of the concept is shown in Fig. 1. Roughly speaking, the security of the system is based on the inability of an adversary (Eve) to distinguish between two symmetrical cases ($R_A=R_0$, $R_B=R_1$ and $R_A=R_1$, $R_B=R_0$) using only *passive* measurements.

We start by noting that the analysis given in [1] for the voltage and current noise density spectra, contains a basic flaw. It completely ignores the finite propagation time between the sender (Alice) and the receiver (Bob) and the finite resistance of the wire connecting them. When the analysis is carried out taking into account the, inevitable, time delay and the resulting transients, or the impedance of the wire, we find that the system becomes vulnerable to eavesdropping, thus invalidating the basic premise of [1]. The analysis leading to the stated conclusion follows.

Referring to Fig. 1, we assume that the wave impedance and length of the transmission line (TL) connecting Alice and Bob are given by $Z_0$ and $L$ respectively. For simplicity, we assume the transmission line is dispersion-less. The voltage and current along the transmission line are given by a superposition of forward and backward propagating waves [2]:

$$V(l,t) = V^+(l,t) + V^-(l,t) \quad (1)$$
$$I(l,t) = [V^+(l,t) - V^-(l,t)]/Z_0$$

At steady-state, because of the random voltage signal generated by the sources, the forward and backward propagating waves at $(0,t)$ due to, say Alice's source, consist of a (infinite) series of time delayed signals emitted at $t-2n\tau$:

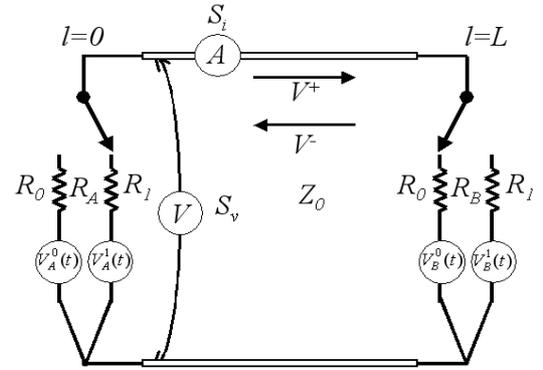

FIG. 1. Schematic of the classical key-distribution system proposed by Kish.

$$V_A^+(0,t) = \frac{Z_0}{R_A + Z_0} \sum_{n=0}^{\infty} (\Gamma_A \Gamma_B)^n V_A(t-2n\tau) \quad (2)$$

$$V_A^-(0,t) = -\frac{Z_0}{(R_A + Z_0)\Gamma_A} V_A(t) + \frac{V_A^+(0,t)}{\Gamma_A}$$

where $V_A(t)$ is the random signal generated by the source at Alice's end, $\tau$ is the propagation time along the TL, and $\Gamma_A$ and $\Gamma_B$ are respectively the reflection coefficients at Alice and Bob's ends defined as:

$$\Gamma_j = \frac{R_j - Z_0}{R_j + Z_0}; \quad j = A, B \quad (3)$$

When one of the parties (say, Alice) switches the resistor (and source) on her side, the abrupt change in the boundary conditions (BC) generates a voltage (and current) wave which propagates toward Bob. If Eve measures the noise spectral density at an asymmetric point on the TL (e.g., close to Alice's end), she can detect this voltage wave and infer Alice's bit.

For simplicity, we pick a specific scenario in which for $t<0$, both Alice and Bob have $R_0$ terminate their end of the TL. At $t=0$ Alice switches $R_1$ on (see Fig. 2). The analysis of the other possibilities is essentially identical leading qualitatively to similar conclusions. We divide the analysis into two cases: 1) The signal propagation time along the line is much longer than the correlation time of the noise generators (or the Johnson noise), i.e., the system is a *distributed* system. 2) The signal propagation time along the line is much shorter than the correlation time of the noise generators, i.e., the system is a *lumped* system. The analysis described by Kish [1] is restricted to the second case and, as shown in the following analysis, this restriction is crucial because it practically eliminates the possibility of utilizing wide bandwidth noise source such as the Johnson noise.

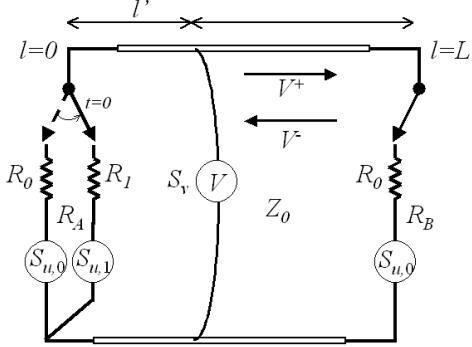

FIG. 2. Determining the exchanged bit by transient analysis of the transmission line.

**2. Case 1 – sources with short correlation time**

In this section we assume that $<V_A(t)V_A(t')> = \delta_{tt'} \cdot W \cdot R_A$ where $W$ is the scaling factor connecting the impedance of the resistor and the variance of the corresponding noise source (for Johnson noise $W$ is given by $4kT$). Because the sources $V_A$ and $V_B$ are independent, the overall noise spectral density measured by Eve is the sum of the separate contribution of each of them. Using (2) and (3) we find that the noise voltage spectral density measured at $l=0$ (i.e. close to Alice's end) for $t<0$ due to Alice's source is given by:

$$S_{u,A}(l=0) = \frac{WR_0 Z_0^2}{(R_0+Z_0)^2} + \frac{W(R_0-Z_0)^4}{8Z_0(Z_0^2+R_0^2)} \quad (4)$$

where in order to sum the infinite series we used the assumption that the signal correlation time is very short. It should be noted that the end points of the TL (i.e. $l=0$, $l=L$), are unique because some of the terms of $V^+$ and $V^-$ add *coherently*. At an arbitrary point $l'$ along the TL, the forward and backward waves due to Alice's source are given by:

$$V_A^+(l',t) = \frac{Z_0}{R_A+Z_0}\sum_{n=0}^{\infty}(\Gamma_A\Gamma_B)^n V_A(t-\tau'-2n\tau)$$

$$V_A^-(l',t) = \frac{Z_0}{(R_A-Z_0)}\left\{-V_A(t+\tau') + \sum_{n=0}^{\infty}(\Gamma_A\Gamma_B)^n V_A(t+\tau'-2n\tau)\right\} \quad (5)$$

where $\tau'$ is the signal propagation time from $l=0$ to $l=l'$. Except for the middle point of the TL ($\tau'=\tau/2$), the terms in (5) add incoherently and the noise spectral density generated by Alice's source at $l'$ is given by:

$$S_{u,A}(l') = \frac{WR_A Z_0(Z_0^2+R_B^2)}{2(Z_0^2+R_A R_B)(R_A+R_B)} \quad (6)$$

The contribution of Bob's source at $l'$ for is evaluated similarly yielding a similar expression where $R_A$ and $R_B$ are interchanged. Note, that for the symmetric case $R_A=R_B$, the overall noise spectral density measured at $l'$ is given by $Z_0 W/2$, i.e., *completely independent* of the resistors value. This outcome demonstrates explicitly the importance of propagation effects along the TL. In contrast with Kish's result (eq. (4) in [1]), in the symmetric case the adversary cannot gain knowledge of the bit values selected by Alice and Bob.

At a first glance, the last conclusion seems to allow Alice and Bob to increase their key-establishing rate by a factor of two because secure communication can take place even in the symmetric case. However, when Alice and/or Bob switch their resistors (and sources), the change creates electromagnetic waves that travel towards Alice and Bob and can be detected by Eve and used to determine the exchanged bit.

Returning to the specific example, at $t<0$, the voltage spectral density measured by Eve at some point along the line is:

$$S_u(l') = \tfrac{1}{2}WZ_0 \quad (7)$$

When Alice switches $R_1$ on, at time $t$, the abrupt change in the BC generates a voltage (and current) wave which propagates toward Bob at a velocity of $V_p=L/\tau$. This wave consists of two contributions: 1) The new noise source, associated with $R_1$, connected by Alice and 2) A change in the reflection coefficient of the left propagating wave ($V^-$) at Alice's end. Since the signals generated by Alice and Bob's noise sources are not correlated, we can calculate the contribution of each source separately and sum them to obtain the power density spectra. The voltage measured by Eve at $l=l'$ and $t>\tau'$ due to Alice's source is, therefore:

$$V(l',t) = \frac{1}{Z_0+R_1}\Big\{Z_0 V_1^A(t-\tau') + (Z_0-R_1)V^-(0,t-\tau')\Big\} + V^-(l',t) \quad (8)$$

Note, that the three terms in (8) are mutually incoherent. Again, for simplicity, we analyze only the first pass of these waves in the transmission line (which is the most dominant one), showing that Eve can learn of the exchanged bit by detecting the change of the voltage generated by these waves. From (8), the noise spectral density due to Alice's source is:

$$S_{u,A}(l') = \frac{WZ_0}{(R_1+Z_0)^2}\left\{R_1 Z_0 + \frac{(R_0-Z_0)^2(Z_0^2+R_1^2)}{4(Z_0^2+R_0^2)}\right\} \quad (9)$$

It should be emphasized that (9) holds only for $\tau'<t<2\tau-\tau'$, i.e., before the reflection of the emitted signal from Alice's (new) source from Bob's end reaches $l=l'$. The contribution of Bob's source to the noise spectral density at $l=l'$ is due to the change in the reflection coefficient at $l=0$ and is given by:

$$S_{u,B}(l') = \frac{WR_1 Z_0(R_0+Z_0)^2}{4(Z_0^2+R_0^2)(R_1+Z_0)} \quad (10)$$

The sum of (9) and (10) yields the over all noise spectral density measured by Eve:

$$S_u(l') = \tfrac{1}{2}WZ_0 + \frac{WZ_0}{2(R_1+Z_0)^2}\left\{R_1^2 + \frac{R_1 Z_0^2(3R_1+Z_0)}{2(Z_0^2+R_0^2)}\right\} \quad (11)$$

Comparing (11) to (7) we find that at $t=\tau$ ($\tau$ after Alice switched $R_1$ on), the overall voltage noise spectral density measured by Eve changes by a quantity equal to the second term in the RHS of (11). In addition, because the contributions of $R_0$ and $R_1$ to that term are not symmetric, Eve can determine whether Alice switched her resistor from $R_0$ to $R_1$ or vice versa. Thus, by monitoring the temporal evolution of the noise density or $<V^2>$ at two points along the TL (one closer to Alice and the other closer to Bob), Eve can determine which resistors (and sources) were selected by Alice and Bob and gain complete knowledge of the exchanged bit.

## 3. Case 2 – sources with long correlation time

In this section we assume, as in [1], that the bandwidth of the noise sources is narrow, i.e., the voltage of the sources does not vary much during the propagation time $\tau$. Under this assumption, the steady-state analysis in [1] is accurate because the system is practically a lump system. However, when Alice and/or Bob switch their resistors, the assumption of the narrow bandwidth sources (and hence, the lumped circuit approximation) becomes invalid. Assuming the two sources are uncorrelated, i.e., $V_1^{(A)}(t) \neq V_0^{(A)}(t)$ and $V_1^{(B)}(t) \neq V_0^{(B)}(t)$, switching from, say, $R_0$ to $R_1$ generates a voltage discontinuity which propagates in speed $V_p$ towards the other side. Thus, similar to case 1, by monitoring the temporal evolution of the voltage at two points along the TL (one closer to Alice and the other closer to Bob), Eve can determine whether Alice and/or Bob switched their resistors (and sources) and gain complete knowledge of the exchanged bit. Note, that unlike case 1, Eve's measurement cannot reveal whether the switching was from $R_0$ to $R_1$ or vice versa. Nevertheless, when an identical resistors scenario ($R_A=R_B$) occurs, which on average happens with probability of 0.5, Eve can determine the value of the resistors and use this information to evaluate the previous and subsequent key bits.

Finally, we show that Eve can also exploit the finite resistance of the wire connecting Alice and Bob to determine the value of $R_A$ and $R_B$ (and consequently the exchanged bit) even without resorting to temporal analysis.

Referring to Fig. 3, we assume that Eve measures the voltage and current at an asymmetrical point along the wire, i.e., $R_{W1} \neq R_{W2}$. The corresponding voltage and current noise density spectra of Eve measurement are given by:

$$\langle V_E^2 \rangle = \frac{W[R_A \cdot (R_B + R_{W2}) + R_B \cdot (R_A + R_{W1})]}{(R_A + R_B + R_{W2} + R_{W1})} \Delta f \qquad (12)$$

$$\langle I_E^2 \rangle = \frac{W \cdot (R_B + R_A)}{(R_A + R_B + R_{W2} + R_{W1})} \Delta f$$

The current noise spectral density can be used to determine the sum of $R_A$ and $R_B$ which indicates whether they are identical or not. For the relevant case, i.e. $R_A \neq R_B$, Eve can use the voltage noise spectral density to distinguish between the two possibilities ($R_A=R_0$, $R_B=R_1$ or vice versa) and determine the exchanged bit.

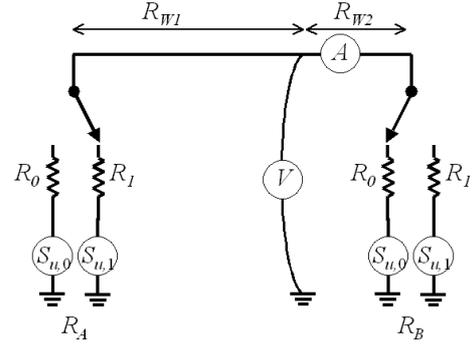

FIG. 3. Determining the exchanged bit using the resistance of the transmission line.

As a concrete example, let us consider a key distribution system employing a 100km long copper wire having a 1mm diameter. The corresponding impedance of this wire is ~2kΩ. This resistance is non-negligible and, depending of Eve's ability to accurately measure the density spectra, it allows her to determine Alice and Bob's selection of resistors.

To conclude, we study the security level provided by the key-distribution scheme suggested by Kish [1]. While at steady state it is impossible to determine the resistors configuration, we show that an adversary can gain complete knowledge of the exchanged bits by using a passive attack strategy exploiting the finiteness of the impedance of the wire connecting Alice and Bob, or the transient response of the system after the resistors have been switched at the end of one (or more) of the parties. The vulnerability of Kish's scheme to the later is crucial because the transient response of the system cannot be eliminated, thus preventing Alice and Bob from obtaining any level of secure key distribution.

Although the specific scheme suggested by Kish turns out, eventually, to be vulnerable to passive attacks, the underlying idea is interesting and worth pursuing. While classical key-distribution systems, which are based on other principles, may not be able to provide *unconditional* security, they may provide *technological* or *practical* security. Unlike QKDS, classical systems do not require single photon sources and detectors, thus allowing secure communication to take place over longer ranges with greater key-establishing rates using currently available components and technologies. Such systems may prove to be both an efficient intermediate solution for secure key-distributions as well as a complementary technology to QKD, especially for long haul links.

The J. S. is grateful to Yuval Cassuto, Barak Dayan and Israel Klich for stimulating discussions and useful comments. A. Y. acknowledges the support of the Defense Advanced Research Projects Agency.

# A Response to Kish comment on "A Classical Key-Distribution System based on Johnson (like) noise – How Secure?" Physics/0602013

Jacob Scheuer
*Center for the Physics of Information, California Inst. of Technology, Pasadena, CA*

Amnon Yariv
*Department of Applied Physics, California Inst. of Technology, Pasadena, CA*

**Abstract**: We have a few short comments regarding Kish's response to our paper "A Classical Key-Distribution System based on Johnson (like) noise – How Secure?" [1].
*Keywords*: Secure communication; Classical information; Eavesdropper detection; Electronic noise; Quantum key distribution.

A recent preprint published by Kish [2] included a response to our criticism of the "Kirchoff-Loop-Johnson-Noise" system [1]. We have a few short comments that will conclude our involvement in this issue.

1) The response [2] does not address numerically or analytically <u>any</u> of points we raised.
2) Kish claims that "…the Scheuer-Yariv manuscript does not identify any security holes in the idealized (mathematical) case of the KLJN cipher". Indeed our main point is that the neglect of wire resistance, propagation delay and the coherence time of noise-like signals renders the mathematical model useless as an approximation of reality. We have quantified our arguments. We will appreciate a numerical/analytical response as an appropriate to a scientific discussion.
3) We completely agree with Kish [2] that "…only generic comments have been published and a thorough analysis of the practical security design aspects the KLJN cipher is still missing…".